# PRICING DERIVATIVES IN HERMITE MARKETS

**Svetlozar T. Rachev**
*Texas Tech University and GlimmAnalytics*

**Stefan Mittnik**
*Ludwig Maximilians University of Munich*

**Frank J. Fabozzi**
*EDHEC Business School*

*Zari's corrections: All comments of Stefan are taken care of in the text. My changes are in red. December 24, 2014, 6 am*

**Abstract:** We introduce Hermite fractional financial markets, where market uncertainties are described by multidimensional Hermite motions. Hermite markets include as particular cases financial markets driven by multivariate fractional Brownian motion and multivariate Rosenblatt motion. Conditions for no-arbitrage and market completeness for Hermite markets are established. Perpetual derivatives, perpetual bonds, forwards, and futures are priced. The corresponding partial and partial-differential equations are derived.

**Keywords**: Hermite processes; fractional Brownian motion; Rosenblatt processes: no-arbitrage and completeness for fractional markets; derivative pricing; term-structure of interest rates; forward and future contracts

# 1.INTRODUCTION

We introduce a novel method for pricing derivatives in fractional markets. All existing fractional market models[1] assume that the riskless asset has the same dynamics as in the classical option pricing models developed by Black and Scholes (1973) and Merton (1973), hereafter referred to as the BSM model. But that assumption leads to the existence of arbitrage trading strategies (see Rogers (1997) and Shiryayev (1998)). Mixing fractional Brownian motion (FBM) with Brownian motion (BM)[2] changes the setting completely as now the overall market driver is no longer a fractional process. Applying Wick integration, Hu and Øksendal (2003) show that an FBM market has no arbitrage opportunities. However, Wick integrations has no economic intuition (see, for example, Björk and Hult (2005)) and the corresponding replicating strategies are very restrictive.[3] In this paper, we propose a different approach. We postulate that in arbitrage-free complete fractional markets, exhibiting long-range dependence (LRD), the riskless asset, if publicly traded,[4]

---

[1] See, for example, Sottinen (2001), Bender, Sottinen, Valkeila (2007), Mishura (2008), and Rostek (2009).

[2] See Mishura (2008) , Kozachenko, Melnikov and Mishura (2014).

[3] In the FBM-market setup of Elliot and Van der Hoek (2003), the simple buy-and-hold strategy is not self-financing (which is indeed discouraging).

[4] In this paper, we do not discuss the existence of a riskless asset. This has been subject of considerable debate recently with Fisher (2013) arguing: "The idea of risk-free sovereign bonds

should have a cumulative rate which increases in time as a regularly varying function (RVF)[5] of order *greater that one*. This is a necessary for no-arbitrage in fractional markets. Indeed, in the BSM-model the variances of the increments of the BM increase linearly in the increments' durations. In sharp contrast, in fractional markets, such as an FBM-market, the variance of the increments of the market driver increases to infinity as RVF of order greater than one[6]. In fractional markets exhibiting LRD, the stock traders are clairvoyant.[7] Thus, the deterministic cumulative return process of the riskless asset should grow faster in time than the linear growth of the riskless-asset return process in the BSM-model. Suppose that in a fractal market exhibiting LRD, a trader can trade a riskless asset which has the same dynamics as the riskless asset in the BSM-model. The trader also can trade a risky asset whose fractal-LRD price dynamics exhibit a stochastic trend with a mean path above the deterministic path of the riskless asset price. Then, the trader will

---

is best thought of as an oxymoron or as an anomaly of recent history. It is not a useful, necessary or an enduring feature of the financial landscape."

[5] See Seneta (1976).

[6] In our setting, a monotone transformation of the time (a market timing transformation, or a time-subordination) can transform the dynamics of the riskless asset in the Hermite market to become identical to the dynamics of the riskless asset in the BSM market.

[7] This is because the risky assets (the stocks) are driven by a fractional process with LRD and the increments of the process are positively correlated.

easily realize various arbitrage trades.[8] This observation is at the heart of our method outlined below.

The paper is organized as follows. In Section 2 we introduce Hermite markets, in which the market uncertainty is modeled by Hermit motion.[9] Our choice for using Hermite markets to model a general fractional market is motivated by the flexibility[10] of Hermit motion as a self-similar process with stationary increments with rich autocorrelation dynamics. In Section 3 we establish the conditions for guaranteeing that a Hermite market is arbitrage-free and complete. The valuation of perpetual derivatives, perpetual bonds, forward and future contracts in a Hermite market is derived in Section 4. Concluding remarks are in Section 5. The proofs are given in the Appendix.

## 2.THE PRIMITIVES OF A HERMIT MARKET

In this section we state the main properties of a Hermit motion market.

### 2.1 **The Hermite Motion as a Model For Market Uncertainty**

---

[8] Examples of such arbitrage strategies are well-known, see Shiryayev (1998) and Rostek (2009).

[9] The parameters of Hermit motions $\mathcal{H}^{(H,\mathscr{k})}(t), t \geq 0, H \in \left(\frac{1}{2}, 1\right)$ are $\in \mathbb{N}, H \in \left(\frac{1}{2}, 1\right)$. $H$ is the Hurst index, also known as the self-similarity index, see Torres and Tudor (2009). We denote $\mathbb{N} = \{1,2, \dots\}$. For $\mathscr{k} = 1$, HM is FBM. For $\mathscr{k} > 1, \mathcal{H}^{(H,\mathscr{k})}$ is a non-Gaussian process, and $\mathcal{H}^{(H,2)}$ is known as Rosenblatt motion, see Taqqu (2011).

[10] This flexibility will be advantageous fitting models empirically.

The **Hermite Motion** (HM),[11] $\mathcal{H}^{(H,k)}(t), t \geq 0, k \in \mathbb{N}, H \in (\frac{1}{2}, 1)$ is defined by

(2.1) $$\mathcal{H}^{(H,k)}(t) = C^{(H,k)} \int_{\mathcal{D}^k} K_t^{(H,k)}\left(v^{(1)}, \dots, v^{(k)}\right) dB\left(v^{(1)}\right) \dots dB\left(v^{(k)}\right),$$

$t \geq 0$[12], where

$(\boldsymbol{HMi})$ $\mathcal{D}^k \coloneqq \left\{\mathbb{v} = \left(v^{(1)}, \dots, v^{(k)}\right) \in R^k : v^{(i)} \neq v^{(j)}, i, j = 1, \dots, k, i \neq j\right\}$;

$(\boldsymbol{HMii})$ For a given $t \geq 0$, the kernel $K_t^{(H,k)}(\mathbb{v}), t \geq 0, \mathbb{v} = \left(v^{(1)}, \dots, v^{(k)}\right) \in R^k$, is defined by[13]

(2.2) $$K_t^{(H,k)}(\mathbb{v}) \coloneqq \int_0^t \left[\prod_{j=1}^{k} \left(s - v^{(j)}\right)_+^{\frac{H-1}{k} - \frac{1}{2}}\right] ds$$

---

[11] See Taqqu (1979), Dobrushin (1979), Dobrushin and Major P. (1979) , Dehling and Taqqu (1989), Lacey (1991), Lavancier F.(2006) , Maejima and Tudor (2007),Tudor (2008),Torres and Tudor (2009), Pipiras and Taqqu (2010), Chronopoulou, Tudor and Viens (2011), Tudor (2013), Marty (2013), Bai and Taqqu (2014), Sun and Cheng (2014), Clausel et al (2014), and Fauth and Tudor (2016).

[12] The integral is understood as Winer-Itô multiple integral, cf. Dobrushin (1979), Nualart (2006), and Clausel et al (2014).

[13] $a_+^b := \begin{cases} a^b, & if\ a \geq 0 \\ 0, & if\ a < 0 \end{cases}, b \in R$. For every given $t \geq 0$, the kernel $K_t^{(H,k)}(\mathbb{v})$, is symmetric and has a finite $\mathcal{L}_2\left(R^k\right)$-norm: $\left\|K_t^{(H,k)}\right\|_{\mathcal{L}_2\left(R^k\right)} = \sqrt{\int_{R^k} \left(K_t^{(H,k)}(\mathbb{v})\right)^2 d\mathbb{v}} < \infty$ . Thus, $\mathcal{H}^{(H,k)}(t), t \geq 0$ is well-defined process.

$(\boldsymbol{HMiii})$ $H \in \left(\frac{1}{2}, 1\right)$ is the Hurst index (index of self-similarity);[14]

$(\boldsymbol{HMiv})$ $C^{(H,\ell)} > 0$ is a normalizing constant such that $\mathbb{E}\left(\mathcal{H}^{(H,\ell)}(1)\right)^2 = 1$.[15]

$(\boldsymbol{HMv})$ $B(v), v \in R$ is a two-sided BM[16]defined on $(\Omega, \mathcal{F}, \{\mathcal{F}_t\}_{t\geq 0}, \mathbb{P})$.

An alternative representation,[17] $\mathcal{H}^{(H,\ell)}(t), t \geq 0$, is given by

(2.3) $$\mathcal{H}^{(H,\ell)}(t) = c^{(H,\ell)} \int_{R^{\ell}} \frac{e^{it\sum_{j=1}^{\ell} u^{(j)}} - 1}{i\left[\sum_{j=1}^{\ell} u^{(j)}\right]\left|\prod_{j=1}^{\ell} u^{(j)}\right|^{\frac{2H-2+\ell}{2\ell}}} B^{(\mathbb{C})}\left(du^{(1)}\right) \dots B^{(\mathbb{C})}\left(du^{(\ell)}\right),$$

where $c^{(H,\ell)}$ is a normalizing constant, so that $\mathbb{E}\left(\mathcal{H}^{(H,\ell)}(1)\right)^2 = 1$, and $B^{(\mathbb{C})}(du)$ is a complex random measure generated by a standard Brownian motion.

---

[14] See, for example, Samorodnitsky (2016) for an extensive study of LRD processes.

[15] $C^{(H,\ell)} = \left(\sqrt{\ell!}\left\|K_1^{(H,\ell)}(\mathbb{v})\right\|_{\mathcal{L}_2(R^{\ell})}\right)^{-1}$, $C^{(H,1)} = \sqrt{\frac{2H\Gamma\left(\frac{3}{2}-H\right)}{\Gamma\left(\frac{1}{2}+H\right)\Gamma(2-2H)}}$ and $C^{(H,2)} = \frac{\Gamma\left(1+\frac{H}{2}\right)\sqrt{\frac{H}{2}(2H-1)}}{\Gamma(\frac{H}{2})\Gamma(1-H)}$, see Clausel et al (2014).

[16] The two-sided Brownian motion $B(v), v \in R$ is defined as follows:

$$B(v) = \begin{cases} B^{(1)}(v), for\ v \geq 0 \\ B^{(2)}(-v), for\ v < 0 \end{cases},$$

where $B^{(1)}(t), t \geq 0$ and $B^{(2)}(t), t \geq 0$ are two independent standard Brownian motions.

[17] See Taqqu (1979).

The existence of the HM follows from the non-central invariance principle:[18]

$$\frac{1}{H}\sum_{j=1}^{\lfloor nt\rfloor}\mathbb{g}^{(\mathcal{k})}\left(\xi^{(j)}\right) \to^{(weakly)} \left(\mathcal{H}^{(H,\mathcal{k})}(t)\right)_{0\le t\le T},$$

where $(i)$ $\xi^{(j)}, j = 0, \pm 1, \pm 2 \dots$, is a stationary Gaussian sequence, with $\mathbb{E}\xi^{(j)} = 0$, $\mathbb{E}\left(\xi^{(j)}\right)^2 = 1$ and covariance function $\rho^{(\xi)}(n) = \mathbb{E}\left[\xi^{(0)}\xi^{(n)}\right]$ having power decay for some slowly varying function[19] $L(n), n = 1,2, \dots$,

$$\lim_{n\uparrow\infty}\frac{\rho^{(\xi)}(n)}{L(n)n^{\frac{2H-2}{\mathcal{k}}}} < \infty;$$

$(ii)$ $\mathbb{g}^{(\mathcal{k})}: R \to R$, where $\mathbb{E}\mathbb{g}^{(\mathcal{k})}\left(\xi^{(0)}\right) = 0, \mathbb{E}\left(\mathbb{g}^{(\mathcal{k})}\left(\xi^{(0)}\right)\right)^2 < \infty$, and $\mathbb{g}^{(\mathcal{k})}$ has Hermite rank $\mathcal{k}$.[20]

The basic properties of the HM, $\mathcal{H}^{(H,\mathcal{k})} = \left\{\mathcal{H}^{(H,\mathcal{k})}(t), t \ge 0\right\}$, are:

---

[18] See Dobrushin and Major (1979), Taqqu (1979), and Torres and Tudor (2009).

[19] Seneta (1976).

[20] Let $g^{(m)}(x) = (-1)^m e^{\frac{x^2}{2}}\frac{d}{dx}e^{-\frac{x^2}{2}}, x \in R,$ be a Hermite polynomial of order $m = 0,1, \dots$ The function $\mathbb{g}^{(\mathcal{k})}: R \to R$ has Hermite rank $\mathcal{k}$, if $\mathbb{g}^{(\mathcal{k})}$ has the following representation:

$$\mathbb{g}^{(\mathcal{k})} = \sum_{m\ge 0} c^{(m)}\, g^{(m)}(x), c^{(m)} \coloneqq \frac{1}{m!}\mathbb{E}\left\{\mathbb{g}^{(\mathcal{k})}\left(\xi^{(0)} g^{(m)}\left(\xi^{(0)}\right)\right)\right\},$$

with $\mathcal{k} = \min\left\{m: c^{(m)} \ne 0\right\}$, see Torres and Tudor (2009).

$\boldsymbol{\mathcal{HBP}^{(1)}}$: $\mathcal{H}^{(H,1)}$ is a FBM and, thus, a Gaussian process. For every $k \geq 2$, $\mathcal{H}^{(H,k)}$[21] is a non-Gaussian process. For all $k \in \mathbb{N}$, $\mathcal{H}^{(H,k)}$ is neither a semimartingale nor a Markov process.

$\boldsymbol{\mathcal{HBP}^{(2)}}$: For all $k \in \mathbb{N}, H \in \left(\frac{1}{2}, 1\right)$, $\mathcal{H}^{(H,k)}$ trajectories are nowhere differentiable. However, for every $0 < \alpha < H$, $\mathcal{H}^{(H,k)}$ trajectories are $\alpha$-Hölder continuous ($\alpha$-Lipchitz continuous), that is, for some $C(H,k) > 0$:

$$\left|\mathcal{H}^{(H,k)}(t) - \mathcal{H}^{(H,k)}(s)\right| \leq C(H,k)|t-s|^{\alpha}, \text{ for all } t \geq 0, s \geq 0.$$

$\boldsymbol{\mathcal{HBP}^{(3)}}$: The trajectories of $\mathcal{H}^{(H,k)}$ are of bounded *Φ- variation*,[22] where

$$\Phi(\mathrm{u}) = \frac{\mathrm{u}^{1/\mathrm{H}}}{\left[\log\left(\log\left(\frac{1}{\mathrm{u}}\right)\right)\right]^{\frac{k}{2H}}}, u > 0.$$ [23]

$\boldsymbol{\mathcal{HBP}^{(4)}}$: Let $V^{(N)}(\mathcal{H}^{(H,k)})$ be the *centered quadratic variation* of $\mathcal{H}^{(H,k)}$, i.e.,

(2.5) $$V^{(N)}\left(\mathcal{H}^{(H,k)}\right) = \sum_{n=0}^{N} \left\{ \begin{array}{c} \left(\mathcal{H}^{(H,k)}\left(t^{(n+1)}\right) - \mathcal{H}^{(H,k)}\left(t^{(n)}\right)\right)^2 - \\ -\mathbb{E}\left[\left(\mathcal{H}^{(H,k)}\left(t^{(n+1)}\right) - \mathcal{H}^{(H,k)}\left(t^{(n)}\right)\right)^2\right] \end{array} \right\},$$

---

[21] $\mathcal{H}^{(H,2)}$ is known as the Rosenblatt process, see Taqqu (2011).

[22] The Φ- variation of a function $f: [0,1] \to R$ is defined by

$V^{(\Phi)}(f) \coloneqq limsup_{\{\mathcal{P}^{(n)}, d(\mathcal{P}^{(n)}) \to 0, n\uparrow\infty\}} \sum_{k=1,\dots,n,(t^{(0)},t^{(1)},\dots,t^{(n)}) \in \mathcal{P}^{(n)}} \Phi(\left|f\left(t^{(k)}\right) - f(t^{(k-1)})\right|)$,

where $\mathcal{P}^{(n)} \coloneqq \left\{\left(t^{(0)}, t^{(1)}, \dots, t^{(n)}\right): 0 = t^{(0)} < t^{(1)} < \cdots < t^{(n)} = 1\right\}$ *and* $d\left(\mathcal{P}^{(n)}\right) \coloneqq max_{k=1,\dots,n}\left|t^{(k)} - t^{(k-1)}\right|$.

[23] Basse-O'Connor and Weber (2016).

where $t^{(n)} - t^{(n-1)} = \gamma^{(N)} > 0, t^{(0)} = 0$; and let

(2.6) $$\delta^{(N)}\left(\mathcal{H}^{(H,\hslash)}\right) \coloneqq \sqrt{\mathbb{E}\left[\left(V^{(N)}(\mathcal{H}^{(H,\hslash)})\right)^2\right]}.$$

Then,[24]

**(I)** if $\hslash = 1$ and $\frac{1}{2} < H \le \frac{3}{4}$, then the limit $lim_{N\uparrow\infty} \frac{\delta^{(N)}\left(\mathcal{H}^{(H,\hslash)}\right)}{\left(\gamma^{(N)}\right)^{2H}\sqrt{N}} < \infty$ exists and is finite, and

(2.7) $$\frac{V^{(N)}\left(\mathcal{H}^{(H,\hslash)}\right)}{\delta^{(N)}\left(\mathcal{H}^{(H,\hslash)}\right)} \Longrightarrow^{distr} \boldsymbol{\mathcal{N}}(0,1);$$

**(II)** if $\hslash = 1$ and $\frac{3}{4} < H < 1$, then the limit $lim_{N\uparrow\infty} \frac{\delta^{(N)}\left(\mathcal{H}^{(H,\hslash)}\right)}{\left(\gamma^{(N)}\right)^{2H} N^{2H-1} logN} < \infty$ exists and is finite, and

(2.8) $$\frac{V^{(N)}\left(\mathcal{H}^{(H,\hslash)}\right)}{\delta^{(N)}\left(\mathcal{H}^{(H,\hslash)}\right)} \Longrightarrow^{distr} \left(\boldsymbol{\mathcal{R}}^{(H)}(0,1)\right)^{1-\frac{2(1-H)}{\hslash}};$$

**(II)** if $\hslash > 1, H \in \left(\frac{1}{2}, 1\right)$, then the limit $lim_{N\uparrow\infty} \frac{\delta^{(N)}\left(\mathcal{H}^{(H,\hslash)}\right)}{\left(\gamma^{(N)}\right)^{2H} N^{\left(1-\frac{2(1-H)}{\hslash}\right)}} < \infty$ exists and is finite, and (2.8) holds.

$\boldsymbol{\mathcal{H}BP}^{(5)}$: $\mathcal{H}^{(H,\hslash)}$ has stationary increments, zero mean, variance $\mathbb{E}\left[\left(\mathcal{H}^{(H,\hslash)}(t)\right)^2\right] = t^{2H}$, and the covariance function $\rho^{(H,\hslash)}$ is given by

---

[24] See Clausel et al. (2016). In what follows: (i) $\Longrightarrow^{distr} \boldsymbol{\mathcal{N}}(0,1)$ is a standard normal random variable (rv); (ii) $\mathcal{R}(0,1)$ is a standard Rosenblatt variable, that is, $\boldsymbol{\mathcal{R}}^{(H)}(0,1)$ has the same distribution as $\mathcal{H}^{(H,2)}(1)$.

(2.9) $\rho^{(H,k)}(t,s) = \mathbb{E}\left(\mathcal{H}^{(H,k)}(t)\mathcal{H}^{(H,k)}(s)\right) = \frac{1}{2}(t^{2H} + s^{2H} - |t-s|^{2H}\}), s \geq 0, t \geq 0;$

$\boldsymbol{\mathcal{H}BP^{(6)}(Long - range\ dependence\ (LRD))}$: Let $\Delta^{(H,k)}(n) \coloneqq \mathcal{H}^{(H,k)}(n+1) - \mathcal{H}^{(H,k)}(n)$, $n = 0,1,\ldots,$ be the sequence of unit increments of $\mathcal{H}^{(H,k)}$. Then $lim_{n\uparrow\infty} n^{2-2H}\mathbb{E}\left(\Delta^{(H,k)}(n)\Delta^{(H,k)}(0)\right) = H(2H-1)$, and, in particular,

(2.10) $$\sum_{n=1}^{\infty} \mathbb{E}\left(\Delta^{(H,k)}(n)\Delta^{(H,k)}(0)\right) = \infty.$$

$\boldsymbol{\mathcal{H}BP^{(7)}}$: $\mathcal{H}^{(H,k)}$ is a self-similar process with Hurst index $H$,

(2.11) $$\mathcal{H}^{(H,k)}(ct) \triangleq c^H\mathcal{H}^{(H,k)}(t), t \geq 0, c > 0.$$

The smoothness of $\mathcal{H}^{(H,k)'}s$ trajectories allows us to define stochastic integrals with respect to $\mathcal{H}^{(H,k)}$ in a pathwise sense.[25] Stochastic calculus with HM is based on the *fractional Stratonovich integral*$(SI)$: For a continuous function, $f: [0,T] \to R$, the *Stratonovich Integral (SI)* is denoted by $\int_0^T f(s) \circledast d\mathcal{H}^{(H,k)}(t)$ and defined as a limit of the Riemann sums[26]

$$\int_0^T f(s) \circledast d\mathcal{H}^{(H,k)} = lim_{n\uparrow\infty} \sum_{k=0,,t^{(k)}=\frac{k}{n}T,k=0,\ldots,n}^{n-1} f\left(t^{(k)}\right)\left(\mathcal{H}^{(H,k)}(t^{(k+1)}) - \mathcal{H}^{(H,k)}(t^{(k)})\right).$$

---

[25] See, for example, Russo and Vallois (1993) and Neuenkirch and Nourdin (2007).

[26] Because $H \in \left(\frac{1}{2},1\right)$, the integral $\int_0^T f(s) \circledast \mathcal{H}^{(H,k)}(t) =$

$$= lim_{n\uparrow\infty} \sum_{k=0,,t^{(k)}=\frac{k}{n}T,k=0,\ldots,n}^{n-1} f\left((1-\delta)t^{(k)} + \delta t^{(k+1)}\right)\left(\mathcal{H}^{(H,k)}(t^{(k+1)}) - \mathcal{H}^{(H,k)}(t^{(k)})\right)$$

has the same value for all $\delta \in [0,1]$, see Duncan (2000).

This implies the following chain-rule: given a sufficiently smooth function $G(x,t), x \in R, t \geq 0$,

(2.12) $$G\left(\mathcal{H}^{(H,\text{k})}(t+s), t+s\right) = G\left(\mathcal{H}^{(H,\text{k})}(t), t\right) +$$

$$= \int_t^{t+s} \frac{\partial G\left(\mathcal{H}^{(H,\text{k})}(u), u\right)}{\partial x} * d\mathcal{H}^{(H,\text{k})}(t) + \int_t^{t+s} \frac{\partial G\left(\mathcal{H}^{(H,\text{k})}(u), u\right)}{\partial u} du,$$

or in differential terms:[27]

$$dG\left(\mathcal{H}^{(H,\text{k})}(t), t\right) = \frac{\partial G(\mathcal{H}^{(H,\text{k})}(t),t)}{\partial x} * d\mathcal{H}^{(H,\text{k})}(t) + \frac{\partial G\left(\mathcal{H}^{(H,\text{k})}(t),t\right)}{\partial t} dt.$$

## 2.2. The Dynamic of the Riskless Asset in a Hermite Market

Let $\mathfrak{L}^{(H,\text{k})}$ be the space of all functions $\mu^{(\circ)}: [0,\infty) \times R^{\text{k}} \to R$, having a representation of the form:

(2. 13) $$\mu^{(\circ)}(t, \mathbb{v}) = \mu(t) K_t^{(H,\text{k})}(\mathbb{v}),$$

where $K_t^{(H,\text{k})}(\mathbb{v})$ is the kernel defined in (2.2), and $\mu(t), t \geq 0,$ is a continuous and uniformly bounded real-valued function, designated as a **basic rate function** (BRF).

Define the *cumulative Hermite rate*, $\mu^{(cum,H,\text{k})}(t), t \geq 0,$ *generated by* $\mu^{(\circ)} \in \mathfrak{L}^{(H,\text{k})}$ as follows:

(2.14) $$\mu^{(cum,H,\text{k})}(t) := C^{(H,\text{k})} \int_{R^{\text{k}}} \left[K_t^{(H,\text{k})}(\mathbb{v}) \mu^{(\circ)}(t,\mathbb{v})\right] d\mathbb{v}$$

$$= C^{(H,\text{k})} \mu(t) \int_{R^{\text{k}}} \left[K_t^{(H,\text{k})}(\mathbb{v})\right]^2 d\mathbb{v} < \infty$$

---

[27] In the literature, two alternative notations are used instead of $\ldots * dB_u^{(H)}$: $(i)$ $\ldots.\delta dB_u^{(H)}$ and $(ii)$ $(S)$ $\ldots.dB_u^{(H)}$.

Note that $\mathbb{K}(t) = \int_{R^{\Bbbk}} \left[K_t^{(H,\Bbbk)}(\mathbb{v})\right]^2 d\mathbb{v} = \left\|K_1^{(H,\Bbbk)}(\mathbb{v})\right\|_{\mathcal{L}_2(R^{\Bbbk})}^2 < \infty$ is homogeneous of order $2H$ and, thus, $\mathbb{K}(t) = \mathbb{K}(1)t^{2H}$. Denoting $D^{(H,\Bbbk)} = C^{(H,\Bbbk)}\mathbb{K}(1) = \frac{1}{\sqrt{\Bbbk!}}\left\|K_1^{(H,\Bbbk)}(\mathbb{v})\right\|_{\mathcal{L}_2(R^{\Bbbk})}$, we have

(2.14) $$\mu^{(cum,H,\Bbbk)}(t) = D^{(H,\Bbbk)}\mu(t)t^{2H}.$$

Define the *(instantaneous) Hermite rate*, $\mu^{(H,\Bbbk)}(t), t \geq 0,$ *generated by the BRF* $\mu(t), t \geq 0,$ as follows:

(2.15) $$\mu^{(H,\Bbbk)}(t) = \frac{\partial}{\partial t}\mu^{(cum,H,\Bbbk)}(t).$$

If $\mu(s) = \mu$ is a constant, then the instantaneous mean return dynamics is given by $\mu^{(H,\Bbbk)}(t) = \mu D^{(H,\Bbbk,)}t^{2H-1}, t \geq 0$, which is a concave increasing function for $t \geq 0$. This is in sharp contrast to the BSM model, where the instantaneous mean return dynamic is constant.

Let $\mathrm{r}^{(0)}: [0,\infty)\times R^{\Bbbk} \to R$ belong to $\mathfrak{L}^{(H,\Bbbk)}$, and

(2.16) $$r^{(\circ)}(t,\mathbb{v}) = r(t)K_t^{(H,\Bbbk)}(\mathbb{v})$$

for some BRF $r(s), s \geq 0,$ such that $lim_{s\geq 0}\left\{r(s) + \frac{1}{r(s)}\right\} < \infty$. Let $r^{(cum,H,\Bbbk)}(t), t \geq 0,$ be *cumulative Hermite rate generated by* $r^{(\circ)}(s,\mathbb{v})$:

(2.17) $$r^{(cum,H,\Bbbk)}(t) := C^{(H,\Bbbk)} \int_{R^{\Bbbk}} \left[K_t^{(H,\Bbbk)}(\mathbb{v})r^{(\circ)}(t,\mathbb{v})\right] d\mathbb{v} = D^{(H,\Bbbk)}r(t)t^{2H}.$$

Assume that there is a publicly traded asset (designated as the **Hermite riskless asset** $\mathcal{M}^{(H,\Bbbk)}$) whose unit price is

(2.18) $$M^{(H,\Bbbk)}(t) = \exp\left\{r^{(cum,H,\Bbbk)}(t)\right\}.$$

We call $r^{(H,\Bbbk)}(t) := \frac{\partial}{\partial t}r^{(cum,H,\Bbbk)}(t), t \geq 0,$ the **Hermite instantaneous riskless rate.**

### 2.3 The Dynamic of the Risky Assets in a Hermite Market

Let $\mathbb{H}^{(H,\mathscr{k})}(t) = \left(\mathcal{H}^{(H,\mathscr{k},1)}(t), \dots, \mathcal{H}^{(H,\mathscr{k},d)}(t)\right), t \geq 0$ be a vector of $d$-independent HM's,

$$\mathcal{H}^{(H,\mathscr{k},m)}(t) = C^{(H,\mathscr{k})} \int_{\mathcal{D}^{\mathscr{k}}} K_t^{(H,\mathscr{k})}\left(v^{(1)}, \dots, v^{(\mathscr{k})}\right) dB^{(m)}\left(v^{(1)}\right) \dots dB^{(m)}\left(v^{(\mathscr{k})}\right),$$

$t \geq 0, m = 1, \dots, d,$ defined on a stochastic basis $(\Omega, \mathcal{F}, \{\mathcal{F}_t\}_{t \geq 0}, \mathbb{P})$ generated by the vector $\mathbb{B}(t) = (B^{(1)}(v), \dots, B^{(d)}(v))$ of $d$ −independend two-sided Brownian motions, $B^{(k)}(v), v \in R, k = 1, \dots, d.$

There are $d$ primary risky assets designated as $\mathfrak{S} = \left(\mathbb{S}^{(1)}, \dots, \mathbb{S}^{(d)}\right)$ with fractal price dynamics per asset unit (a share): $\mathcal{S}(t) = \left(S^{(1)}(t), \dots, S^{(d)}(t)\right), t \geq 0,$

$$\frac{dS^{(j)}(t)}{S^{(j)}(t)} = \mu^{(H,\mathscr{k},j)}(t)dt + \sum_{m=1}^{d} \sigma^{(j,m)}(t) \circledast d\mathcal{H}^{(H,\mathscr{k},m)}(t),$$

where *instantaneous drift functions* $\mu^{(H,\mathscr{k},j)}(t), t \geq 0, j = 1, \dots, d$ are defined by

$$\mu^{(H,\mathscr{k},j)}(t) = \frac{\partial}{\partial t} \mu^{(cum,H,\mathscr{k},j)}(t)$$

with $\mu^{(cum,H,\mathscr{k},j)}(t) \coloneqq C^{(H,\mathscr{k})} \int_{R^{\mathscr{k}}} \left[K_t^{(H,\mathscr{k})}(\mathbb{v})\mu^{(j)}(t,\mathbb{v})\right] d\mathbb{v} = D^{(H,\mathscr{k})}\mu^{(j)}(t)t^{2H}$, for some BRF $\mu^{(j)}(t) > 0$, and

(2.19) $$\mu^{(j)}(t,\mathbb{v}) \coloneqq \mu^{(j)}(t) K_t^{(H,\mathscr{k})}(\mathbb{v}).$$

The *volatility functions* $\sigma^{(j,k)}(t), t \geq 0,\ j = 1, \dots, d, k = 1, \dots, d,$ are assumed to be continuous and uniformly bonded.

Pathwise integration leads to the explicit representation for the stock prices given by

$S^{(j)}(t) = S^{(j)}(0)\exp\{\mu^{(cum,H,\mathscr{k},j)}(t) + \sum_{m=1}^{d} \sigma^{(j,m)}(t)\,\mathcal{H}^{(H,\mathscr{k},m)}(t)\}, t \geq 0, j = 1, \dots d.$

## 3. HERMIT MARKETS: SUFFICIENT CONDITIONS FOR NO-ARBITRAGE AND COMPLETENESS

Consider a trader (designated as ב) investing in a Hermit market $\mathfrak{M} = (\mathcal{M}^{(0)}, \dots, \mathcal{M}^{(d)}) := (\mathcal{M}^{(H,\mathscr{k})}, \mathbb{S}^{(1)}, \dots, \mathbb{S}^{(d)})$, where $\mathcal{M}^{(H,\mathscr{k})}$ is the riskless asset and $\mathbb{S}^{(1)}, \dots, \mathbb{S}^{(d)}$ are the risky assets. Denote by $\mathbb{X}(t), t \geq 0$, the vector of the corresponding price processes, i.e., $\mathbb{X}(t) = \left(X^{(0)}(t), X^{(1)}(t), \dots, X^{(d)}(t)\right) = \left(M^{(H,\mathscr{k})}(t), S^{(1)}(t), \dots, S^{(d)}(t)\right), t \geq 0.$

Let ב′ have initial capital (wealth) $w(0) \in R$ and investment horizon $\mathcal{T} \leq \infty$, and let $w(t), t \geq 0,$ denote the wealth at time $t$. It is assumed that ב trades without additional inflow or outflow of funds. At time $t \geq 0$, ב is investing the entire wealth $w(t)$ in a portfolio with value $P(t)$: $w(t) = P(t)$. ב employs a trading strategy $\Theta(t) = \left(\theta^{(0)}(t), \dots, \theta^{(d)}(t)\right), t \geq 0$. The strategy $\Theta$ is an $\{\mathcal{F}_t\}_{t\geq 0}$- adapted bounded continuous process, where $\theta^{(j)}(t) \in R, j = 0, \dots d,$ is the number of units ב bought from security $\mathcal{M}^{(j)}$ at time $t \geq 0$. At any time $t \geq 0$, $P(t) = \sum_{j=0}^{d} \theta^{(j)}(t)\, X^{(j)}(t)$. Then $G^{(\Theta,\mathbb{X})}(t) := \sum_{j=0}^{d} \int_0^t \theta^{(j)}(s) dX^{(j)}(s), t \geq 0$ is ב′$s$ gain process from trading strategy $\Theta(t) = \left(\theta^{(0)}(t), \dots, \theta^{(d)}(t)\right)$. Strategy $\Theta(t), t \geq 0$ is self-financing if $P(t) = P(0) + G^{(\Theta,\mathbb{X})}(t)$.

A **self-financing strategy is an arbitrage** if either $(i)$ $P(0) < 0$, but for some $T > 0$, $\mathbb{P}(P(T) \geq 0) = 1$, or $(ii)$ $P(0) \leq 0$, but for some $T > 0$, $\mathbb{P}(P(T) \geq 0) = 1$ and $\mathbb{P}(P(T) > 0) > 0$. . Consider as **market-deflator,** $Y(t), t \geq 0$, the **fractional discount factor** $Y(t) =$

$\frac{1}{M^{(H,k)}(t)}$. Then, the price dynamics of the deflated market $\mathbb{X}^{(Y)}(t) =$

$\left(X^{(0)^{(Y)}}(t), X^{(1)^{(Y)}}(t), \dots, X^{(d)^{(Y)}}(t)\right) = \left(1, \frac{S^{(1)}(t)}{M^{(H,k)}(t)}, \dots, \frac{S^{(d)}(t)}{M^{(H,k)}(t)}\right), t \geq 0,$ are given by

(3.1) $$d\frac{S^{(j)}(t)}{M^{(H,k)}(t)} = S^{(j)}(t)d\frac{1}{M^{(H,k)}(t)} + \frac{1}{M^{(H,k)}(t)}dS^{(j)}(t) =$$

$$= \frac{S^{(j)}(t)}{M^{(H,k)}(t)}\left(\left(\mu^{(H,k,j)}(t) - r^{(H,k)}(t)\right)dt + \sum_{m=1}^{d}\sigma^{(j,m)} \circledast d\mathcal{H}^{(H,k,m)}(t)\right)$$

Let $(\Omega, \mathcal{F}, \{\mathcal{F}_t\}_{t\geq 0}, \mathbb{Q}) \sim (\Omega, \mathcal{F}, \{\mathcal{F}_t\}_{t\geq 0}, \mathbb{P})$ be such that the following conditions hold:

$(\boldsymbol{EMM^{(1)}})$ $(\Omega, \mathcal{F}, \{\mathcal{F}_t\}_{t\geq 0}, \mathbb{Q})$ is generated by $d-$dimensional BM $\mathbb{B}^{(\mathbb{Q})}(v) = \left(B^{(1,\mathbb{Q})}(v), \dots, B^{(d,\mathbb{Q})}(v)\right), s \in R$, where $B^{(m,\mathbb{Q})}(v), v \in R, m = 1, \dots, d$, are independent two-sided Brownian motions;

$(\boldsymbol{EMM^{(2)}})$ on the natural world $\mathbb{P}$,

(3.2) $$dB^{(k,\mathbb{Q})}\left(v^{(1)}\right) \dots dB^{(k,\mathbb{Q})}\left(v^{(k)}\right) = dB^{(k)}\left(v^{(1)}\right) \dots dB^{(k)}\left(v^{(k)}\right) + \mathfrak{z}^{(k)}\left(v^{(1)}, \dots, v^{(k)}\right).$$

The vector $\mathfrak{Z}(\mathbb{v}) = \left(\mathfrak{z}^{(1)}(\mathbb{v}), \dots, \mathfrak{z}^{(d)}(\mathbb{v})\right)^T, \mathbb{v} \in R^{k}$, (representing the **market price of risk)** satisfies the linear system for all $\mathbb{v} \in R^d, t \geq 0,$[28]

(3.3) $$\mu^{(j)}(t, \mathbb{v}) - r^{(\circ)}(t, \mathbb{v}) = \sum_{m=1}^{d}\sigma^{(j,m)}(t)\mathfrak{z}^{(m)}(\mathbb{v}), j = 1, \dots, d,$$

where $\mu^{(j)}(t, \mathbb{v}) = \mu^{(j)}(t)K_t^{(H,k)}(\mathbb{v})$ (see (2.19)) and $r^{(\circ)}(t, \mathbb{v}) = r(t)K_t^{(H,k)}(\mathbb{v})$ (see (2.16)).

---

[28] The market price of risk $\mathfrak{z}^{(m)}(\mathbb{v})$ should satisfy (3.3) for all $t \geq 0$. If $\mu(t) = \mu, r(t) = r$, (3.3) is the standard no-arbitrage condition for the BSM market.

We assume that $\mathfrak{Z}(\mathbb{v}) = \left(\mathfrak{z}^{(1)}(\mathbb{v}), \dots, \mathfrak{z}^{(d)}(\mathbb{v})\right)'$ exists and is unique, implying that the market $(\mathcal{M}^{(H,\ell)}, \mathbb{S}^{(1)}, \dots, \mathbb{S}^{(d)})$ is characterized by no arbitrage opportunities and completeness[29]

Note that $\frac{S^{(j)}(t)}{\mathcal{M}^{(H,\ell)}(t)}, j = 1, \dots, d,$ is *not* a $\mathbb{Q}$ -martingale. To circumvent that obstacle when valuating derivatives in the Hermite market, we use replicating self-financing strategies (see Section 4) to find derivative dynamics.

## 4. BSM PERPETUAL OPTION PRICING IN FRACTAL MARKETS

Suppose in the market $\mathfrak{M} = \left(\mathcal{M}^{(0)}, \dots, \mathcal{M}^{(d)}\right) \coloneqq (\mathcal{M}, \mathbb{S}^{(1)}, \dots, \mathbb{S}^{(d)})$, the risky assets $\mathbb{S}^{(1)}, \dots, \mathbb{S}^{(d)}$ pay dividends at Hermite dividend rates $\boldsymbol{\delta}^{(\boldsymbol{H})}(t) = \left(\delta^{(H,1)}(t), \dots, \delta^{(H,d)}(t)\right)$, where

$\delta^{(H,\ell,j)}(t) \coloneqq \frac{\partial \delta^{(cum,H,\ell,j)}(t)}{\partial t}, t \geq 0$, with

(4.1) $$\delta^{(cum,H,\ell)}(t) := C^{(H,\ell)} \int_{R^{\ell}} \left[K_t^{(H,\ell)}(\mathbb{v}) \delta^{(\circ)}(t, \mathbb{v})\right] d\mathbb{v} = D^{(H,\ell)} \delta(t) t^{2H},$$

for some $\delta^{(\circ)} \in \mathfrak{L}^{(H,\ell)}, \delta^{(\circ)}(t, \mathbb{v}) \coloneqq \delta(t) K_t^{(H,\ell)}(\mathbb{v}), j = 1, \dots, d.$ The price dynamics of the stock $\mathbb{S}^{(j)}, j = 1, \dots, d,$ are given by

(4.2) $$\frac{dS^{(j)}(t)}{S^{(j)}(t)} = (\mu^{(H,\ell,j)}(t) - \delta^{(H,\ell,j)}(t))\, dt + \sum_{m=1}^{d} \sigma^{(j,m)} \divideontimes d\mathcal{H}^{(H,\ell,m)}(t).$$

### 4.1. <u>**Valuation of Perpetual Derivatives in Fractal Market With Dividends**</u>

---

[29] We use arguments very similar to those given in Duffie 2001, Chapter 6, to prove this result.

Consider a perpetual derivative with price process $\mathrm{G_t} = g\left(t, S^{(1)}(t), \dots, S^{(d)}(t)\right), t \geq 0$, where $g(\boldsymbol{t}, \mathbb{x}), \mathbb{x} = \left(x^{(1)}, \dots, x^{(d)}\right),\ t \geq 0, x^{(j)} > 0, t \geq 0, j = 1, \dots, d$ is sufficiently smooth function.

*PROPOSITION: Suppose (4.2) holds. Then the ECC- price process* $G_t = \left(t, S^{(1)}(t), \dots, S^{(d)}(t)\right), t \geq 0$ *satisfies the following partial differential equation:*

*(4.3)* $$\frac{\partial g(t,\mathbb{x})}{\partial t} + \sum_{j=1}^{d} \frac{\partial g(t,\mathbb{x})}{\partial x^{(j)}} x^{(j)} \left(r^{(H,\mathscr{k})}(t) - \delta^{(H,\mathscr{k},j)}(t)\right) - r^{(H,\mathscr{k})}(t) g(t, \mathbb{x}) = 0.$$

*Proof: See Appendix A1.*

Let $\beta: [0, \infty) \to R, \boldsymbol{\alpha} = \left(\alpha^{(1)}, \dots, \alpha^{(d)}\right) \in R^d$, and let

$$\mathrm{G}_{\mathrm{t}}^{(\beta,\boldsymbol{\alpha})} \coloneqq g\left(t, S^{(1)}(t), \dots, S^{(d)}(t)\right) = M^{(H,\mathscr{k})}(t)^{\beta(t)} \prod_{j=1}^{d} \left[\left(S^{(j)}(t)\right)^{\alpha^{(j)}}\right], t \geq 0.$$

Then security $\mathfrak{E}^{(\beta,\boldsymbol{\alpha})}$ with price process $\mathrm{G}_{\mathrm{t}}^{(\beta,\boldsymbol{\alpha})}, t \geq 0$, can be publicly traded together with the assets $(\mathcal{M}, \mathbb{S}^{(1)}, \dots, \mathbb{S}^{(d)})$ if and only if

(4.3) $$\frac{\partial\left[\beta(t) r^{(cum,H,\mathscr{k})}(t)\right]}{\partial t} + \sum_{j=1}^{d} \alpha^{(j)} \left(r^{(H,\mathscr{k})}(t) - \delta^{(H,\mathscr{k},j)}(t)\right) - r^{(H,\mathscr{k})}(t) = 0.$$

for all $t \geq 0$.[30] That is, at any time $t \geq 0$, trader ℶ can buy $\mathfrak{E}^{(\beta,\boldsymbol{\alpha})}$ for $\mathrm{G}_{\mathrm{t}}^{(\beta,\boldsymbol{\alpha})}$ and sell it for $\mathrm{G}_{\mathrm{T}}^{(\beta,\boldsymbol{\alpha})}$ at any time $T > t$.

#### 4.1.1 *Alternative Valuation BSM Formula for Hermite Markets*

---

[30] That is, the extended market $\left(\mathfrak{E}^{(\beta,\boldsymbol{\alpha})}, \mathcal{M}, \mathbb{S}^{(1)}, \dots, \mathbb{S}^{(d)}\right)$ is arbitrage-free and complete.

The Hermite market consists of two assets (a) the $H$-*riskless* asset, $\mathcal{M}^{(H,r)}$, with price dynamics

$$M^{(H)}(t) = M^{(H)}(0)e^{rt^{2H}}, t \geq 0, M^{(H)}(0) > 0, H \in \left(\frac{1}{2}, 1\right], r > 0;$$

and (b) the $H$-*risky asset*, $\mathcal{S}^{(H,\mu,\sigma)}$, with price dynamics

$$S^{(H)}(t) = S^{(H)}(0)e^{\mu t^{2H} + \sigma \mathcal{H}^{(H,\hslash)}(t)}, t \geq 0, S^{(H)}(0) > 0, \hslash \in \mathbb{N}, \mu > r > 0, \sigma > 0.$$

Consider an ECC with price dynamics $G(t) = g(S^{(H)}(t), t)$, where $g(x,t), x \geq 0, t \geq 0$, is sufficiently smooth. Assume that ב is taking a short position in the ECC-contract and is hedging her risk with $\Delta(t)$ shares of the stock and $b(t)$ units of the bond. At time $t$, $ב's$ self-financing portfolio is $P(t) = -G(t) + \Delta(t)S^{(H)}(t) + b(t)M^{(H)}(t)$ . ב choses $\Delta(t)$ and $b(t)$ so that $P(t + dt) = 0$, which results in $P(t) = 0$, and $dP(t) = -dg\left(S^{(H)}(t), t\right) + \Delta(t)dS^{(H)}(t) + b(t)dM^{(H)}(t) = 0$. Thus, pathwise integration leads to $\Delta(t) = \frac{\partial g\left(S^{(H)}(t),t\right)}{\partial x}$ and $b(t)2Ht^{2H-1}rM^{(H)}(t) = \frac{\partial g\left(S^{(H)}(t),t\right)}{\partial t}$. Thus, $g(S^{(H)}(t), t) = \frac{\partial g\left(S^{(H)}(t),t\right)}{\partial x} S^{(H)}(t) +$ $+ \frac{\partial g\left(S^{(H)}(t),t\right)}{\partial t} \frac{1}{2Ht^{2H-1}r}$. That is, $g(x,t)$ satisfies the following partial differential equation,

$$\frac{\partial g(x,t)}{\partial t} + r2Ht^{2H-1}\frac{\partial g(x,t)}{\partial x}x - r2Ht^{2H-1}g(x,t) = 0.$$

which is, indeed, a special case of (4.3).

Consider a perpetual derivative $\mathcal{D}^{(\alpha,\beta)}$, with price process $g\left(S^{(H)}(t), t\right) = S^{(H)}(t)^{\alpha}M^{(H)}(t)^{\beta}, \alpha \in R, \beta \in R$. Then, for $\mathcal{D}^{(\alpha,\beta)}$ to be publicly traded (together with $\mathcal{M}^{(H,r)}$ and $\mathcal{S}^{(H,\mu,\sigma)}$ ) it is necessary and sufficient that $\alpha + \beta = 1$.That is, in the publicly traded Hermite market $(\mathcal{M}^{(H,r)}, \mathcal{S}^{(H,\mu,\sigma)})$ any perpetual derivative $\mathcal{D}^{(\alpha)} = \mathcal{D}^{(\alpha,1-\alpha)}, \alpha \in R$, with price process $G^{(\alpha)}(t) := S^{(H)}(t)^{\alpha} M^{(H)}(t)^{1-\alpha}, t \geq 0$, can be also publicly traded. In other words, at any time $t \geq 0$, ב can buy $\mathcal{D}^{(\alpha)}$ for $\$G^{(\alpha)}(t)$, and sell it for $\$G^{(\alpha)}(T)$, at time $T > t$. Furthermore, any asset with price process $G^{(\gamma)}(t) := \int_R S^{(H)}(t)^{\alpha(u)} M^{(H)}(t)^{1-\alpha(u)} \gamma(u) du$ [31] can be publicly traded in the Hermite market $\left(\mathcal{M}^{(H,r)}, \mathcal{S}^{(H,\mu,\sigma)}\right)$.

### 4.1.2 *Hermite Market with Subordinated Time Dynamics*

Consider again the Hermite market $(\mathcal{M}^{(H,r)}, \mathcal{S}^{(H,\mu,\sigma)})$ and assume that the **physical time** $(\boldsymbol{t \geq 0})$ is subordinated in the following manner: $\tau^{2H} = t$. We can view $\boldsymbol{\tau \geq 0}$ **as the market-time**. Assume that securities $\mathcal{M}^{(H,r)}, \mathcal{S}^{(H,\mu,\sigma)}$ have the following dynamics with respect to the market time: $M^{(H)}(\tau) = M^{(H)}(0) e^{r\tau^{2H}}$, and $S^{(H)}(\tau) = S^{(H)}(0) e^{\mu\tau^{2H} + \sigma\mathcal{H}^{(H,\mathscr{k})}(\tau)}, \tau \geq 0$. Then, the dynamics of the two assets with respect to the physical time are given by:

$$M^{(H)}(t) = M^{(H)}(0) e^{rt}, M^{(H)}(0),\ S^{(H)}(t) = S^{(H)}(0) e^{\mu t + \sigma\mathfrak{H}^{(H,\mathscr{k})}(t))}, S^{(H)}(0) > 0,$$

where

$$\mathfrak{H}^{(H,\mathscr{k})}(t) := \mathcal{H}^{(H,\mathscr{k})}\left(t^{\frac{1}{2H}}\right) = C^{(H,\mathscr{k})} \int_{\mathcal{D}^{\mathscr{k}}} \mathcal{Z}_t^{(H,\mathscr{k})}\left(v^{(1)}, \dots, v^{(\mathscr{k})}\right) dB\left(v^{(1)}\right) \dots dB\left(v^{(\mathscr{k})}\right),$$

and $\mathcal{Z}_t^{(H,\mathscr{k})}(\mathbb{v}) \coloneqq K_{t^{\frac{1}{2H}}}^{(H,\mathscr{k})}(\mathbb{v})$. Note that $\mathbb{Z}(t) = \int_{R^{\mathscr{k}}} \left[\mathcal{Z}_t^{(H,\mathscr{k})}(\mathbb{v})\right]^2 d\mathbb{v} < \infty$, is homogeneous of order 1 and, thus, $\mathbb{Z}(t) = \mathbb{Z}(1)t$. The **subordinated Hermit process** $\mathfrak{H}^{(H,\mathscr{k})}(t), t \geq 0$, is fractional (self-similar) process with index of self-similarity equal to ½ , but has no stationary increments.

---

[31] The functions $\alpha(u), \gamma(u), u \in R$, are chosen so that $G^{(\gamma)}(t), t \geq 0$, is well defined.

### 4.2. **Hermite Perpetual Zero-Coupon Bond**

Suppose that $\mathfrak{E}^{(\beta,\boldsymbol{\alpha})}$ with price process $\mathrm{G}_t^{(\beta,\boldsymbol{\alpha})}$ is a perpetual bond $\mathrm{G}_t^{(\beta,\boldsymbol{\alpha})} = \mathfrak{b}(t)$,[32] $t \geq 0$. That is, $\mathfrak{E}^{(\beta,\boldsymbol{\alpha})}$ is a perpetual security[33] which bought at time $t \geq 0$, can be sold for $\mathfrak{b}(T), T > t$. Then, setting $\boldsymbol{\alpha} = \mathbf{0}$ in (4.3) results in $\mathfrak{b}(t) = \mathrm{G}_t^{(\beta,\boldsymbol{\alpha})} = M^{(H,\&)}(t)^{\beta(t)} = \exp\{\beta(t) r^{(cum,H,\&)}(t)\}$, $t \geq 0$ where $\frac{\partial[\beta(t) r^{(cum,H,\&)}(t)]}{\partial t} - r^{(H,\&)}(t) = 0$. Thus, $\beta(t) = 1$, is a solution of (4.3)[34], and, and $\mathfrak{b}(t) = M^{(H,\&)}(t) = e^{r^{(cum,H,\&)}(t)} = e^{r^{(cum,H,\&)}(0)} e^{\int_0^t r^{(H,\&)}(s)ds}$.

Consider then a **perpetual Hermite zero-coupon bond** of type $\mathfrak{E}^{(\beta,\boldsymbol{\alpha})}$ (designated as $\mathcal{Z}^{(T)}$) and price process $\Lambda^{((H,\&,T))}(t,T), t \geq 0, \Lambda(T,T) = 1$, and furthermore, *for all* $t \geq 0$,

(4.4) $$\Lambda(t,T) = \frac{M^{(H,\&)}(t)}{M^{(H,\&)}(T)} = e^{-\int_t^T r^{(H,\&)}(s)ds}$$

We call the collection of prices $\mathfrak{L}^{(H,\&)} := \left\{\Lambda(t,T) = e^{-\int_t^T r^{(H,\&)}(s)ds}, t \geq 0, T \geq 0\right\}$, the **Hermite term-structure of interest rates** (HTSIR).

### 4.3 **Hermite Perpetual Forward Contract**

Suppose at time $t \geq 0$, a trader ב is short selling one-share of the risky asset $\mathbb{S} = \mathbb{S}^{(1)}$ which pays no dividends, and simultaneously is using the cash $\$S(t) = \$S^{(1)}(t)$ to buy $S(t)$units of the Hermite riskless asset $\mathcal{M}^{(H,\&)}$. The portfolio is perpetual and can be sold at any time $\tau > t$ for $P^{(t,T)}(\tau)$, where, for all $u \geq t$ , $P^{(t,T)}(u) = -S(u) + S(t)\mathfrak{b}(0)e^{\int_t^u r^{(H,\mathbb{k})}(s)ds}$. We designated as $\mathfrak{F}^{(t)}$ the security with price process $P^{(t,T)}(u), u \geq 0$. Indeed, if $t < u \leq T$,

(4.5) $$P^{(t,T)}(u) = -S(u) + F(t,T)\Lambda(u,T),$$

represents the value of the hedging portfolio at time $u$, replicating a long forward contract with maturity $T > t$ on stock $\mathbb{S}$ and forward (delivery) price $F(t,T) \coloneqq \mathfrak{b}(0)\frac{S(t)}{\Lambda(t,T)}$.[35] Thus, in the trading period $[t,T]$, security $\mathfrak{F}^{(t)}$ can be viewed as long forward contract with maturity $T$, and forward price $F(t,T)$. However, if security $\mathfrak{F}^{(t)}$ is not exercised at some time-instance $u \leq T$ , then at any time $u > T$, $\mathfrak{F}^{(t)}$ still will have a non-zero value and can be publicly traded. That is, why we call $\mathfrak{F}^{(t)}$ a perpetual forward contract.

---

[32] We assume $\mathfrak{b}(t), t \geq 0$, is a deterministic price process.

[33] The security must be of type $\mathfrak{E}^{(\beta,\boldsymbol{\alpha})}$.

[34] Indeed, other solutions of (4.3) are also possible, but we chose the simplest one.

### 4.4 <u>Fractional Futures Contract</u>

Suppose ℶ trades at discrete trading instances $t^{(k,n)} = \frac{k}{n}t, k = 0,1,\dots,n \uparrow \infty$, where $t > 0$ is fixed time-instance. ℶ enters a long futures contract at $t^{(k,n)}$, when the futures price is $\Phi\left(t^{(k,n)}\right)$. At $t^{(k+1,n)}$, when the futures price becomes $\Phi\left(t^{(k+1,n)}\right)$, ℶ receives $\Phi\left(t^{(k+1,n)}\right) - \Phi\left(t^{(k,n)}\right) = \Psi\left(S\left(t^{(k+1,n)}\right), t^{(k+1,n)}\right)$. The underlying asset for the futures contract $\mathbb{S} = \mathbb{S}^{(1)}$, with price process $S(t) = S^{(1)}(t)$,

(4.6) $$\frac{dS(t)}{S(t)} = \mu^{(H,\mathbb{k},S)}(t)\,dt + \sum_{m=1}^{d} \sigma^{(S,m)} \circledast d\mathcal{H}^{(H,\mathbb{k},m)}(t)$$

---

[35] See Appendix A2.

Note that, if $\sigma^{(S)} := \sqrt{\sum_{k=1}^{d} (\sigma^{(S,k)})^2}$, $\mathfrak{B}^{(S)}(t) \coloneqq \sum_{k=1}^{d} \int_0^t \frac{\sigma^{(S,k)}}{\sigma^{(S)}} B^{(k)}(s) ds$ for $t \geq 0$, and $\mathfrak{B}^{(S)}(t) \coloneqq \sum_{k=1}^{d} \int_t^0 \frac{\sigma^{(S,k)}}{\sigma^{(S)}} B^{(k)}(s) ds$, for $t < 0$, then $\mathfrak{B}^{(S)}(t), t \in R$, is a two-sided Brownian motion. Furthermore, $\frac{dS(t)}{S(t)} = \mu^{(H,S)}(t)dt + \sigma^{(S)} \divideontimes d\mathcal{H}^{(H,\Bbbk,S)}(t)$, where

$$\mathcal{H}^{(H,\Bbbk,S)}(t) == C^{(H,\Bbbk)} \int_{\mathcal{D}^{\Bbbk}} K_t^{(H,\Bbbk)}(\mathbb{v})\, d\mathfrak{B}^{(S)}\left(v^{(1)}\right) \dots d\mathfrak{B}^{(S)}\left(v^{(\Bbbk)}\right),$$

The total payoff from the futures contract in $[0,t], \Phi(\mathrm{t}) = \int_0^t \Psi(\mathrm{S(u)}, u)\mathrm{du}$, where $\Psi(\mathrm{x}, t), x > 0, t \geq 0$, is sufficiently smooth. Then, applying a replicating self-financing strategy $a(t) \geq 0, b(t) \geq 0, \Phi(\mathrm{t}) = \mathrm{a(t)S(t) + b(t)M(t)}$, leads to the following partial integro-differential equation (PIDE)[36] for $\Psi(\mathrm{x(t)}, t), t \geq 0$

(4.7) $$\int_0^t \Psi(\mathrm{x(u)}, u)\mathrm{du} = \Psi(\mathrm{x(t)}, t) \frac{\partial\Psi(\mathrm{x(t),t})}{\partial x} + \frac{1}{r^{(H,\Bbbk)}(t)} \Psi(\mathrm{x(t)}, t) \frac{\partial\Psi(\mathrm{x(t)}, t)}{\partial t}.$$

## 5. CONCULSION

We derived conditions for fractal Hermit markets to be arbitrage-free and complete. Because Hermite markets include as special cases the fractional Brownian-motion market and the Rosenblatt-motion market, our results cover those market types as well. Central to our method is the introduction of a riskless asset in a Hermite market with price dynamics faster than the dynamics of the riskless asset in the BSM model. We derive (i) a partial differential equation for the price of perpetual derivative, (ii) a term structure of interest rates in a Hermite market, (iii) the value of a forward contract, and (iv) the partial integro-differential equation for the value of a perpetual futures contract.

[36] The proof is given in Appendix A.3. For numerical solution of PIDE we refer to Appell, Kalitvin and Zabrejko (2000).

# APPENDIX

## A1. PROOF OF PROPOSITION 1

From the path-wise integration formula:

$$dg\left(t, S^{(1)}(t), \dots, S^{(d)}(t)\right) = \frac{\partial g\left(t, S^{(1)}(t), \dots, S^{(d)}(t)\right)}{\partial t} dt +$$

$$+\sum_{j=1}^{d} \frac{\partial g\left(t, S^{(1)}(t), \dots, S^{(d)}(t)\right)}{\partial x^{(j)}} S^{(j)}(t)\left(\mu^{(H,j)}(t) - \delta^{(H,j)}(t)\right) dt +$$

$$+\sum_{j=1}^{d} \frac{\partial g\left(t, S^{(1)}(t), \dots, S^{(d)}(t)\right)}{\partial x^{(j)}} S^{(j)}(t) \sum_{k=1}^{d} \sigma^{(j,k)}(t) \circledast \mathcal{H}^{(H,k)}(t).$$

Consider a self-financing strategy, $g\left(t, S^{(1)}(t), \dots, S^{(d)}(t)\right) = \sum_{j=1}^{d} a^{(j)}(t) S^{(j)}(t) + b(t) M^{(H,k)}$.

Thus, $dg\left(t, S^{(1)}(t), \dots, S^{(d)}(t)\right) = \sum_{j=1}^{d} a^{(j)}(t) dS^{(j)}(t) + b(t) dM^{(H,k)}$. Equating the expressions for $dg\left(t, S^{(1)}(t), \dots, S^{(d)}(t)\right)$ leads to $a^{(j)}(t) = \frac{\partial g\left(M, S^{(1)}(t), \dots, S^{(d)}(t)\right)}{\partial x^{(j)}}, j = 1, \dots, d$ and

$$b(t)M(t) = \frac{1}{r^{(H,\ell)}(t)} \frac{\partial g\left(t, S^{(1)}(t), \dots, S^{(d)}(t)\right)}{\partial t}$$

$$- \frac{1}{r^{(H,\ell)}(t)} \sum_{j=1}^{d} \frac{\partial g\left(t, S^{(1)}(t), \dots, S^{(d)}(t)\right)}{\partial x^{(j)}} S^{(j)}(t) \delta^{(H,j)}(t)$$

Applying $\mathrm{g}\left(\mathrm{t}, \mathrm{S}^{(1)}(\mathrm{t}), \dots, \mathrm{S}^{(\mathrm{d})}(\mathrm{t})\right) = \sum_{\mathrm{j}=1}^{\mathrm{d}} \mathrm{a}^{(\mathrm{j})}(\mathrm{t}) \mathrm{S}^{(\mathrm{j})}(\mathrm{t}) + \mathrm{b}(\mathrm{t}) \mathrm{M}(\mathrm{t})$ results in (4.3).

*Q.E.D.*

## A2. PERPETUAL FORWARD CONTRACT

Suppose at time $t \geq 0$, trader ב forms a portfolio with price process, $P^{(t,T)}(u), u \geq t$ of short position the stock $\mathbb{S}^{(1)}$, and long position of $\frac{S^{(1)}(t)}{\Lambda(t,T)}$ units of $\mathcal{Z}^{(T)}$ −bonds, that is $P^{(t,T)}(t) = -S^{(1)}(t) + \frac{S^{(1)}(t)}{\Lambda(t,T)} \Lambda(t,T) = 0$. Then for $u > t$, the value of ב$'s$ portfolio becomes $P^{(t,T)}(u) = -S^{(1)}(u) + \frac{S^{(1)}(t)}{\Lambda(t,T)} \Lambda(u,T)$. Indeed, if $t < u \leq T$, $P^{(t,T)}(u)$ represents the value of the hedging portfolio at time $u$, replicating a a *long forward contract* with maturity $T > t$ on stock $\mathbb{S}^{(1)}$ paying no dividends. But suppose now that $t > T$. What is the meaning of $P^{(t,T)}(u)$

$= -S^{(1)}(u) + \frac{S^{(1)}(t)}{\Lambda(t,T)} \Lambda(u,T)$ for $u > t > T$? Because $\Lambda(v,T) = e^{-\int_v^T r^{(H)}(s)ds}$, for all $v > 0$,

$P^{(t,T)}(u) = -S^{(1)}(u) + S^{(1)}(t) e^{\int_t^u r^{(H)}(s)ds}$. This price dynamics, corresponds to ב short selling one-share in $\mathbb{S}^{(1)}$ at time $t$ an simultaneously using the cash of \$$S^{(1)}(t)$ to buy $S^{(1)}(t)$ of the fractal riskless asset $\mathcal{M}$. The portfolio is perpetual and can be sold at any time $\tau > t$ for \$$P^{(t,T)}(\tau)$.Q.E.D.

## A3. PERPETUAL FUTURE CONTRACT

By the pathwise integration formula, $d\Phi(\mathrm{t}) = \Psi(\mathrm{S}(\mathrm{t}), t)\left(\frac{\partial\Psi(\mathrm{S}(\mathrm{t}),t)}{\partial x} S(t)\mu^{(S)}(t)\, dt + \frac{\partial\Psi(\mathrm{S}(\mathrm{t}),t)}{\partial t} dt\right) == \Psi(\mathrm{S}(\mathrm{t}), t)\left(\frac{\partial\Psi(\mathrm{S}(\mathrm{t}),t)}{\partial x} S(t)\sigma^{(S)} * Z^{(S)}(t)\right)$. Suppose $\Phi(\mathrm{t}) = \mathrm{a}(\mathrm{t})\mathrm{S}(\mathrm{t}) + \mathrm{b}(\mathrm{t})\mathrm{M}(\mathrm{t})$. Thus, $\mathrm{d}\Phi(\mathrm{t}) = \left(\mu^{(S)}(t)\mathrm{a}(\mathrm{t})\mathrm{S}(\mathrm{t}) + \mathrm{b}(\mathrm{t})\mathrm{M}(\mathrm{t})r^{(H)}(t)\right) dt + \mathrm{a}(\mathrm{t})\mathrm{S}(\mathrm{t})\sigma^{(S)} * Z^{(S)}(t)$

Equating the terms $\mathrm{d}\Phi(\mathrm{t})$ leads to $\mathrm{a}(\mathrm{t}) = \Psi(\mathrm{S}(\mathrm{t}), t)\frac{\partial\Psi(\mathrm{S}(\mathrm{t}))}{\partial x}$, and $\mathrm{b}(\mathrm{t})\mathrm{M}(\mathrm{t}) = \frac{1}{r^{(H)}(t)}\Psi(\mathrm{S}(\mathrm{t}), t)\frac{\partial\Psi(\mathrm{S}(\mathrm{t}),t)}{\partial t}$. Finally, because, $\Phi(\mathrm{t}) = \int_0^t \Psi(\mathrm{S}(\mathrm{u}), u)\mathrm{du} = \mathrm{a}(\mathrm{t})\mathrm{S}(\mathrm{t}) + \mathrm{b}(\mathrm{t})\mathrm{M}(\mathrm{t})$

we have $\int_0^t \Psi(\mathrm{S}(\mathrm{u}), u)\mathrm{du} = \Psi(\mathrm{S}(\mathrm{t}), t)\frac{\partial\Psi(\mathrm{S}(\mathrm{t}))}{\partial x} + \frac{1}{r^{(H)}(t)}\Psi(\mathrm{S}(\mathrm{t}), t)\frac{\partial\Psi(\mathrm{S}(\mathrm{t}),t)}{\partial t}$. That is, $\int_0^t \Psi(\mathrm{x}(\mathrm{u}), u)\mathrm{du} = \Psi(\mathrm{x}(\mathrm{t}), t)\frac{\partial\Psi(\mathrm{x}(\mathrm{t}),\mathrm{t})}{\partial x} + \frac{1}{r^{(H)}(t)}\Psi(\mathrm{x}(\mathrm{t}), t)\frac{\partial\Psi(\mathrm{x}(\mathrm{t}),t)}{\partial t}$. Q.E.D.